\documentclass{article}

\usepackage{microtype}
\usepackage{graphicx}
\usepackage{subfigure}
\usepackage{booktabs} 

\usepackage{hyperref}

\usepackage[accepted]{icml2023-diffxyz}

\usepackage{amsmath}
\usepackage{amssymb}
\usepackage{mathtools}
\usepackage{amsthm}
\usepackage{xspace}

\usepackage[capitalize,noabbrev]{cleveref}

\theoremstyle{plain}

\theoremstyle{definition}

\theoremstyle{remark}

\usepackage[textsize=tiny]{todonotes}

\newcommand{\bd}[1]{\mathbf{#1}}
\newcommand{\bth}{\boldsymbol{\theta}}
\newcommand{\bTh}{\boldsymbol{\Theta}}

\newcommand{\balpha}{\boldsymbol{\alpha}}
\newcommand{\june}{\textsc{June}\xspace}
\newcommand{\gradjune}{\textsc{GradABM-June}\xspace}

\icmltitlerunning{}

\begin{document}

\twocolumn[
\icmltitle{Some challenges of calibrating differentiable agent-based models}

\icmlsetsymbol{equal}{*}

\begin{icmlauthorlist}
\icmlauthor{Arnau Quera-Bofarull}{equal,cs}
\icmlauthor{Joel Dyer}{equal,cs,inet}
\icmlauthor{Anisoara Calinescu}{cs}
\icmlauthor{Michael Wooldridge}{cs}
\end{icmlauthorlist}

\icmlaffiliation{cs}{Department of Computer Science, University of Oxford}
\icmlaffiliation{inet}{Institute for New Economic Thinking, Oxford}

\icmlcorrespondingauthor{Arnau Quera-Bofarull}{arnau.quera-bofarull@cs.ox.ac.uk}
\icmlcorrespondingauthor{Joel Dyer}{joel.dyer@cs.ox.ac.uk}

\icmlkeywords{Machine Learning, ICML}

\vskip 0.3in
]

\printAffiliationsAndNotice{\icmlEqualContribution} 

\begin{abstract}
Agent-based models (ABMs) are a promising approach to modelling and reasoning about complex systems, yet their application in practice is impeded by their complexity, discrete nature, and the difficulty of performing parameter inference and optimisation tasks. This in turn has sparked interest in the construction of differentiable ABMs as a strategy for combatting these difficulties, yet a number of challenges remain. In this paper, we discuss and present experiments that highlight some of these challenges, along with potential solutions.
\end{abstract}

\section{Introduction}
\label{intro}

Agent-based models (ABMs, see \autoref{sec:abm} for a brief overview) have gained considerable popularity across a range of disciplines, due to their ability to accurately simulate complex systems at a granular level. While these models offer unique advantages, their complexity presents significant challenges, for example in terms of parameter calibration \citep[see e.g.][]{dyer2022black, dyer2022calibrating}. For such tasks, multiple factors contribute to their difficulty, including the intractability of the ABM's likelihood function, and the often black-box and non-differentiable nature of the ABM.

These drawbacks of ABMs have motivated research into the construction of differentiable ABMs \citep{gradabm, arnau_quera_bofarull_2023_7623959}, for example through the use of differentiable programming and by exploiting automatic differentiation (AD) frameworks. AD -- a methodological cornerstone in machine learning, largely underpinning the success of deep learning paradigms due to its ability to accurately compute derivatives within models -- circumvents issues present in alternative approaches to model differentiation by applying the chain rule of differentiation at a computational level, resulting in exact derivatives. 

Despite recent progress, the challenges involved in building and benefitting from differentiable ABMs remain under-explored, and there exists little guidance to practitioners interested in implementing and exploiting differentiable ABMs. %
The aim of this paper is therefore to discuss some central challenges in applying AD to ABMs.

\section{Challenges}\label{sec:challenges}

\subsection{Discrete randomness}

The issue of differentiating through discrete structures is inherent in ABMs, which simulate discrete events, transitions, and interactions that are incompatible with conventional AD. Initial efforts to implement AD within ABMs have primarily centred on transforming the ABMs' discrete control flow structure with continuous approximations \cite{andelfingerDifferentiableAgentBasedSimulation2021}. Furthermore, the use of the Gumbel-Softmax (GS) reparametrisation trick \cite{jangCategoricalReparameterizationGumbelSoftmax2017} allows for the differentiation of discrete randomness, and has been deployed effectively in epidemiological ABMs \citep{gradabm}. However, this approach does not provide an ideal solution. While it allows for gradient calculations, GS does not guarantee unbiased or low-variance gradients \cite{huijbenReviewGumbelmaxTrick2022a}. Developing unbiased and lower variance methods such as StochasticAD \cite{aryaAutomaticDifferentiationPrograms2022} in the Julia programming language \cite{bezanson2017julia} is currently an active field of research, but we limit the scope of our discussion here to GS-based methods for discrete ABMs.

Despite the potential lack of robustness of GS, GS-based differentiable ABM implementations have shown great success in improving the calibration \cite{gradabm} and sensitivity analyses \cite{queraDontSimulate} of ABMs. In the Experiments section below, we further show that gradients obtained using the GS trick are robust enough to enable fast and accurate Bayesian inference (see \autoref{sec:june_experiment}).

\subsection{Reverse- vs. Forward-mode AD}
\label{sec:memory}

\begin{figure}
    \centering
    \includegraphics[width=0.9\columnwidth]{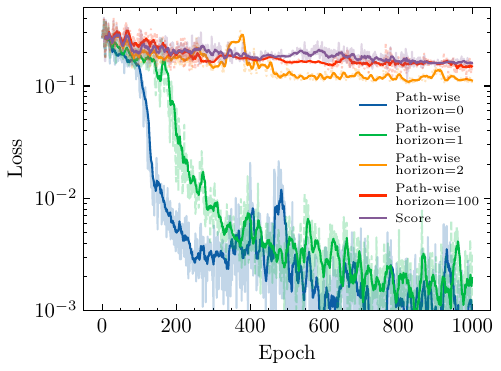}
    \vspace{-1em}
    \caption{Training loss as a function of epochs for each gradient estimation method and gradient horizon value. Solid line shows 10-step moving average; dashed lines are without averaging.}
    \label{fig:bh_horizons}
\end{figure}

In Reverse-mode AD (RMAD), a computation graph must be stored that records all operations performed within the model, such that the gradients of the model outputs with respect to the input parameters can be obtained. This contrasts with Forward-mode AD (FMAD), where the gradients are computed during the forward simulation. There are two important computational considerations when comparing FMAD vs. RMAD. The first is that the computational time associated with FMAD scales with the number of model inputs, while that of RMAD scales with the number of model outputs. In machine learning, the latter option is more prevalent, since machine learning models often have many more inputs than outputs. 
However, the computation graph that RMAD must store in (often GPU) memory can be extremely large, hindering the possibility of differentiating through large models. This is particularly pertinent for ABMs: the size of the computation graph grows with the number of agents and time-steps, which can pose a challenge to the use of RMAD for ABMs with a large number of agents and time-steps.

To address this, in \autoref{sec:june_experiment} we discuss a differentiating strategy that alternates between FMAD and RMAD when calibrating ABMs, and we apply it to an epidemiological simulation involving 8 million agents.%

\subsection{Monte Carlo gradient estimation}

Since ABMs are typically stochastic models, it can often be the case that practitioners are interested in performing an optimisation problem of the form
\begin{equation}\label{eq:gen_obj}
    \min_{\omega \in \Omega} \mathbb{E}_{z \sim p_{\omega}} \left[\mathcal{L}(z)\right],
\end{equation}
where $p_{\omega} \in \{ p_{\omega'} : \omega' \in \Omega \}$ is a probability distribution on some domain $\mathcal{Z}$ indexed by a parameter $\omega$ belonging to some set $\Omega$, and $\mathcal{L} : \mathcal{Z} \to \mathbb{R}$ is a loss function. For example, certain parameter calibration procedures that seek to identify the parameters $\bth$ in some set $\bTh$ that minimise some discrepancy $\mathcal{D}(\cdot, \bd{y})$ between the model output $\bd{x}$ and real-world data $\bd{y}$ can be cast in the form
\begin{equation}
    \min_{\bth \in \bTh} \mathbb{E}_{\bd{x} \sim p(\cdot \mid \bth)}\left[\mathcal{D}(\bd{x}, \bd{y})\right],
\end{equation}
where $p(\cdot \mid \bth)$ is the ABM's likelihood function. The gradients of a differentiable ABM can then be exploited by gradient-assisted methods for minimising the objective in \eqref{eq:gen_obj}, by finding a Monte Carlo estimate of the expression
\begin{equation}\label{eq:grad_gen_obj}
    \nabla_{\omega} \mathbb{E}_{z \sim p_{\omega}}\left[\mathcal{L}(z)\right].
\end{equation}
For differentiable ABMs, a Monte Carlo estimate of this gradient can be obtained using the path-wise derivative via reparametrisation tricks \cite{mohamedMonteCarloGradient}. In such cases, derivatives of the form $\partial \bd{x}_t / \partial \omega_{i}$ will contribute to the estimate. To properly benefit from access to the differentiable ABM's gradients in these settings, it is critical that low-variance, low-bias Monte Carlo estimates of \eqref{eq:grad_gen_obj} are available. However, as we will demonstrate in \autoref{sec:bh}, naively estimating these gradients by accounting for both (a) the explicit dependency of each $\bd{x}_t$ on $\omega_i$, and (b) the implicit dependency on $\omega_i$, mediated by the $\bd{x}_{1:t-1}$ that result from the recursive structure of ABMs, can result in unusable gradient estimates with prohibitively large variances. Consequently, modifications to vanilla AD can become necessary, as illustrated in \autoref{sec:bh}.

\begin{figure*}
    \centering
    \includegraphics[width=1.85\columnwidth]{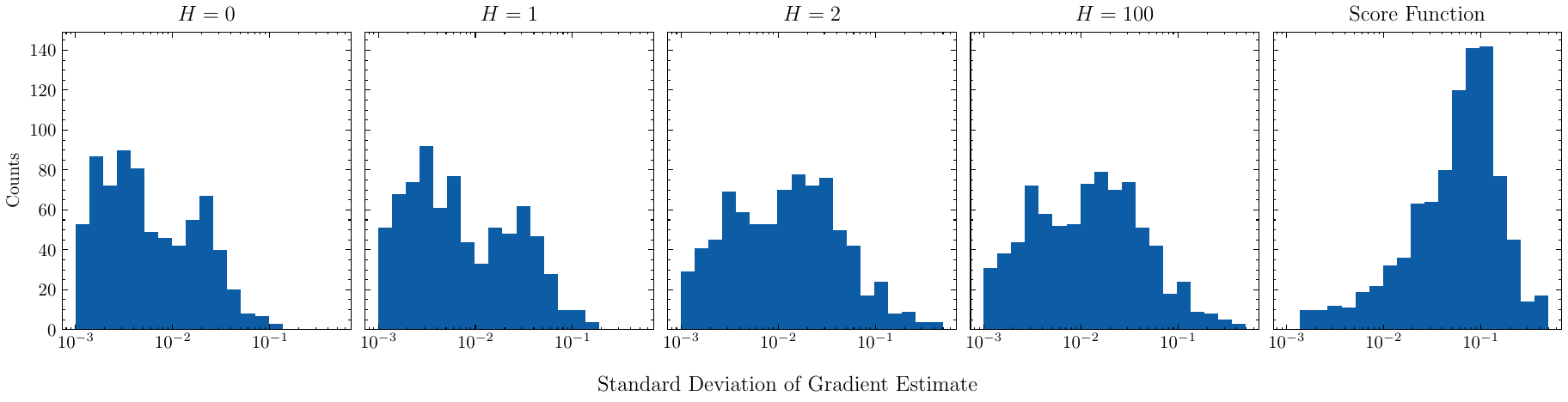}
    \vspace{-1em}
    \caption{Histograms of the standard deviations in estimates of $\partial \mathbb{E}_{q_\phi}\ell(\bd{y}, \bth) / \partial \phi_j$ for different gradient horizons and the score-based method across all $j$ for the Brock \& Hommes model.}
    \label{fig:flow_gradients_stds}
\end{figure*}
\section{Experiments} 

In this section, we present experiments on the use of gradient-assisted calibration methods for two ABMs, where each experiment serves to highlight different combinations of the challenges described in Section \ref{sec:challenges}. 

While there exist %
many different gradient-assisted calibration methods, 
we focus %
on a variational approach to Bayesian parameter inference termed Generalised Variational Inference \citep[GVI,][]{knoblauch2022optimization} -- a likelihood-free Bayesian inference approach that has previously been used to calibrate the parameters $\bth \in \mathbb{R}^d$ of a differentiable ABM \citep{quera2023bayesian}. Here, a variational procedure targets a ``generalised'' posterior \citep{bissiri2016general}
\begin{equation}\label{eq:genpost}
    \pi_{w, \bd{y}}(\bth) \propto e^{-w\cdot \ell(\bd{y}, \bth)} \pi(\bth),
\end{equation}
where $\pi(\bth)$ is a prior distribution, $\ell(\bd{y}, \bth)$ is a loss function capturing the compatibility between the observed data $\bd{y}$ and the behaviour of the ABM at parameter vector $\bth$, and $w > 0$ is a hyperparameter. To target this posterior, we train a normalising flow $q_{\phi}$ with trainable parameters $\phi$ to minimise the Kullback-Liebler divergence $\text{KL}(q_{\phi} \Vert \pi_{w, \bd{y}})$ from $q_{\phi}$ to $\pi_{w, \bd{y}}$, yielding the minimising parameters
\begin{equation}\label{eq:gvi_obj}
    \phi_{w,\bd{y},\pi} 
    = \arg \min_{\phi}\bigg\{ w\,\mathbb{E}_{q_{\phi}}\left[\ell(\bd{y}, \bth)\right] + \text{KL}\left(q_{\phi} \Vert \pi\right)\bigg\}.
\end{equation}
Further details 
are provided in \autoref{app:training}. Code to reproduce the results and perform GVI on differentiable ABMs can be found at \url{https://github.com/arnauqb/blackbirds}.

\subsection{The Brock \& Hommes model}
\label{sec:bh}
The Brock \& Hommes model \cite{brockHeterogeneousBeliefsRoutes1998} is a heterogeneous agent model for the price $\bd{x}_t \in \mathbb{R}$ of an asset over time $1 \leq t \leq T$. At each time step, the agents in the model subscribe to one of a set of $J > 1$ trading strategies, each of which is characterised by a trend-following parameter $g_j$ and bias parameter $b_j$, $j \in \{1, \ldots, J\}$. Following \citet{dyer2022black}, we note that 
the price $\bd{x}_t$ may be written as deterministic transformations $f_t$ of the input parameters $\bth = (g_1, \ldots, g_J, b_1, \ldots, b_J)$, auxiliary parameters $\balpha$, and standard Normal random variables:
\begin{equation}
    \bd{x}_t = f_t(\epsilon_{1}, \ldots, \epsilon_t, \bth, \balpha),\quad \epsilon_t \sim \mathcal{N}(0,1).
\end{equation}
Further details are provided in \autoref{app:BH}. Thus, provided $\ell$ is chosen to be a differentiable function of $\bth$, this enables us to employ gradient-based approaches to minimising the objective \eqref{eq:gvi_obj} that exploit the reparameterisation trick.

Fixing $g_1 = b_1 = b_4 = 0$ and $g_4 = 1.01$, we consider the task of calibrating parameters $g_2, g_3, b_2, b_3$ given synthetic data $\bd{y} = (\bd{y}_1, \ldots, \bd{y}_T)$ generated from the model at $(g_2, g_3, b_2, b_3) = (0.9, 0.9, 0.2, -0.2)$ with $T=100$. We follow \citet{mmdbayes} and 
target the generalised posterior \eqref{eq:genpost} given by the choice
\begin{align}
    \ell(\bd{y}, \bth) &= \text{MMD}^2(\mathbb{P}_T, \mathbb{P}_{\bth}),
\end{align}
where $\text{MMD}^2(\mathbb{P}_T, \mathbb{P}_{\bth})$ is the maximum mean discrepancy between the empirical measure of returns $\mathbb{P}_T = (\bd{y}_1, \ldots, \bd{y}_T)$ and the distribution $\mathbb{P}_{\bth}$ of returns implied by the simulator at parameters $\bth$. Using a Gaussian kernel within the MMD computation, the operations comprising evaluation of $\ell(\bd{y}, \bth)$ are also all differentiable and deterministic, enabling evaluation of the term $\nabla_{\phi} \mathbb{E}_{\bth \sim q_{\phi}} \ell(\bd{y}, \bth)$ in \eqref{eq:gvi_obj} using the reparameterisation trick (see Appendix \ref{app:BH_rep}).

Despite our ability to compute the partial derivatives $\partial \bd{x}_t / \partial \phi_{i}$ exactly, 
the posterior estimator $q_{\phi}$ struggles to train with a gradient-assisted approach to minimising \eqref{eq:gvi_obj}. This can be seen in \autoref{fig:bh_horizons}, in which the objective function decreases slowly with the number of epochs when trained with AdamW \citep{Loshchilov2017DecoupledWD} and using the vanilla pathwise derivative (red curve). Indeed, we see in this case that the access to the simulator's gradients appears to offer no improvement over the score-based gradient, shown with the purple curve and obtained as
\begin{equation}
    \nabla_{\phi} \mathbb{E}_{q_{\phi}}\left[\ell(\bd{y}, \bth)\right] = \mathbb{E}_{q_\phi} \left[\ell(\bd{y}, \bth) \nabla_{\phi} \log q_{\phi}(\bth)\right].
\end{equation}
Drawing inspiration from the literature on backpropagation-through-time (e.g. truncated back-propagation in the context of RNN training, see \citet{sutskever2013training}), we consider pruning a subset of the paths in the computation graph that contribute to each of the $\partial \bd{x}_t / \partial \bth_i$ 
as a possible solution to this problem. We achieve this by invoking an appropriate \texttt{stop\_gradient} operation (e.g. \texttt{.detach()} in \texttt{pytorch}, \citet{pytorch}) on terms $\bd{x}_{t'}$ that (a) contribute directly/explicitly to the evaluation of $\bd{x}_t$ and (b) for which $t' < t - H$ for some ``gradient horizon'' $H \geq 0$. As evident from \autoref{fig:bh_horizons}, we observe that a finite gradient horizon can dramatically improve the gradient-assisted training of $q_{\phi}$. In this experiment, the best performance was observed while using a gradient horizon of $H = 0$.

We posit that this is a manifestation of a bias-variance trade-off in the Monte Carlo gradient estimation step: pruning a subset of paths in the computation graph with the use of a finite gradient horizon may introduce some bias in, but can significantly reduce the variance of, Monte Carlo estimates of the gradient $\nabla_{\phi} \mathbb{E}_{q_{\phi}}\left[\ell(\bd{y}, \bth)\right]$ when employing the pathwise derivative. This hypothesis is supported by \autoref{fig:flow_gradients_stds}, which shows histograms of the standard deviation of the estimates of $\partial \mathbb{E}_{q_{\phi}}\left[\ell(\bd{y}, \bth)\right] / \partial \phi_{j}$ across all $j$ for gradient horizons $H \in \{ 0, 1, 2, 100 \}$, and for the score-based estimator. There, we see that the histogram shifts towards larger values as $H$ increases. Further results supporting this hypothesis are given in Appendix \ref{app:BH_exp}.

\subsection{The JUNE model}
\label{sec:june_experiment}

The \june model \cite{aylett-bullockJuneOpensourceIndividualbased2020} is a large-scale epidemiological ABM of England based on a realistic synthetic population constructed from the English census. Calibrating the original implementation required the construction of a surrogate model due to its high computational cost \cite{vernonBayesianEmulationHistory2022c}. The \gradjune model \cite{arnau_quera_bofarull_2023_7623959} is a differentiable implementation of \june which employs the GS reparameterisation trick to differentiate through discrete randomness. Compared to its non-differentiable counterpart, \gradjune has been used to more efficiently generate parameter point estimates, as well as sensitivity analyses \cite{queraDontSimulate}.

\begin{figure}
    \centering
    \includegraphics[width=0.8\columnwidth]{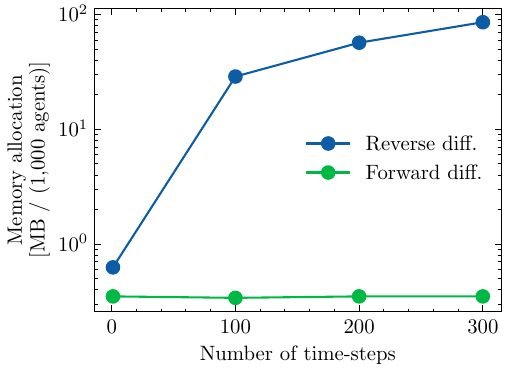}
    \vspace{-1em}
    \caption{Memory consumption of running AD in the JUNE model per 1,000 agents, for the Forward and Reverse mode of AD.}
    \label{fig:june_memory}
\end{figure}
\subsubsection{Reducing memory consumption through forward-mode AD}
\gradjune can simulate the entire English population at a scale of 1:1 --- 53 million people at the time of the 2011 census. As discussed in \autoref{sec:memory}, differentiating through this model using RMAD is challenging due to the high memory demand to store the computation graph. To perform GVI for this model, we implement a hybrid AD technique: we use FMAD to obtain the Jacobian $J_{\bth}$ of the ABM outputs with respect to the ABM parameters, and combine it with RMAD through the flow $q_\phi$, yielding
\begin{align}
   \nabla_{\phi} \mathbb{E}_{q_{\phi}}\left[\ell(\bd{y}, \bth)\right]  &= J_{\bth} (\mathbb{E}_{q_{\phi}}\left[\ell (\mathbf y, \bth)\right]) \cdot \nabla_\phi \bth,\quad \text{with}\\\label{eq:jac_fmad}
    J_{\bth} (\mathbb{E}_{q_{\phi}}\ell (\mathbf y, \bth)) &= \frac{\partial \mathbb{E}_{q_{\phi}}\ell (\mathbf y, \bth)}{\partial \bth} \in \mathbb{R}^{1 \times d}.
\end{align}
Here, \eqref{eq:jac_fmad} is the Jacobian obtained through FMAD and $\nabla_\phi\bth \in \mathbb{R}^{d\times F}$ is the gradient, obtained with RMAD, of the $d$ ABM parameters generated by the normalising flow with parameters $\phi \in \mathbb{R}^F$. 

In \autoref{fig:june_memory}, we plot the memory costs of employing FMAD and RMAD to compute the ABM's Jacobian. We see that the cost of FMAD is independent of the number of time-steps, since no computation graph is stored. In contrast, the cost of RMAD scales linearly with the number of time-steps and agents. Simulating the entire English population for 300 time-steps would require 5TB of memory, while doing so with FMAD would require merely 18GB, regardless of the number of time-steps. The increase in computational time of FMAD comes at an increase of computational cost since it requires $d$ evaluations of the model for $\bth \in \mathbb{R}^d$; however, since these evaluations are embarrassingly parallelisable, the impact on performance can be minimal.

\begin{figure}
    \centering
    \includegraphics[width=0.8\columnwidth]{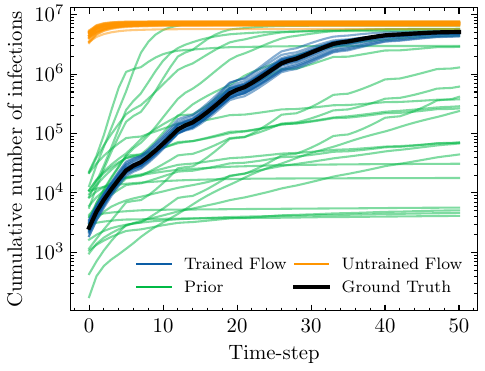}
    \vspace{-1em}
    \caption{Cumulative infections over time. Black: synthetic ground truth; green, orange, blue: runs obtained from samples from the prior, untrained flow, and trained flow, respectively.}
    \label{fig:june_results}
\end{figure}

With the above in mind, we set up an experiment with the London's population (8.1 million people) in \gradjune. We generate a synthetic time-series of daily infections for 50 days using some assumed parameters that we aim to recover through our calibration process. The parameters that we vary are the contact intensities at 10 different locations, as well as the number of initial cases. Further details of the experimental setup are shown in \autoref{app:june}.

\label{sec:june_results}
We apply the GVI procedure to calibrate the \june model with 11 free parameters. The flow converges after approximately 3,000 model evaluations, highlighting the potential for simulation-efficient calibration with gradient-assisted methods. In \autoref{fig:june_results}, we show a comparison of runs obtained by sampling the ABM parameters from the trained flow, the untrained flow, and the prior. This demonstrates that the trained flow generates parameters that result in close agreement between the simulator and ground truth data, while providing useful uncertainty quantification.


\section{Conclusions and discussion}

This study examines some challenges that arise from the application of vanilla AD to ABMs, such as overcoming the inherent discreteness of ABMs and the variance and high computational requirements of passing gradients through large simulators. We have shown that these challenges can be overcome to some extent with different modifications to vanilla AD. As supporting evidence, we successfully calibrate differentiable implementations of the Brock \& Hommes and \june models with these modifications, the latter involving over eight million agents and discrete randomness. In this way, this study helps to pave the way towards robust calibration of large-scale agent-based models. 

\bibliography{main}
\bibliographystyle{icml2023}

\newpage
\appendix
\onecolumn

\section{Agent-based models}\label{sec:abm}

Agent-based modelling is the name given to a broad approach to modelling complex systems that consist of multiple discrete, autonomous, and heterogeneous interacting components -- the ``agents'' of the system. Examples of such complex systems include the housing market \citep{baptista2016staff}, in which a large collection of renters, homeowners, financial institutions etc. interact and take actions which affect, for example, the availability of housing and mortgage rates. An agent-based approach to modelling such a system would model the system at the level of these individual agents in the system, often with the intention of observing how aggregate, macroscopic properties of the system emerge from the microscopic details of the system. 

While this is often a natural approach to modelling systems of this kind, the inherently discrete nature of the model's components and dynamics give rise to difficulties in applying gradient-based optimisation and calibration techniques. We expand on these difficulties in \autoref{sec:challenges}.

\section{The Brock \& Hommes model}\label{app:BH}

The dynamics of the Brock and Hommes model are often expressed as the following system of coupled equations:
\begin{align}
    \bd{x}_{t} &= \frac{1}{R}\left[\sum_{j=1}^{J} \left(g_j \bd{x}_{t-1} + b_j\right)n_{j, t} + \sigma \epsilon_{t}\right],\ \epsilon_{t} \sim \mathcal{N}(0, 1),\\
    n_{j, t} &= \frac{\exp{\left(\beta U_{j,t-1}\right)}}{\sum_{j' = 1}^{J} \exp{\left(\beta U_{j',t-1}\right)}},\\
    U_{j,t-1} &= \left(\bd{x}_{t-1} - R \bd{x}_{t-2}\right)\left(g_j \bd{x}_{t-3} + b_j - R \bd{x}_{t-2}\right),
\end{align}
where $R, \beta, \sigma$ are auxiliary parameters. We fix $J=4, R = 1.01, \sigma = 0.04, g_1 = b_1 = b_4 = 0$, $g_4 = 1.01$, and $\beta = 120$ for the experiment presented in the main body of the paper. 
By rewriting the above system of equations, we are able to find the transition density for observation $\bd{x}_{t+1}$ as
\begin{equation}\label{eq:BH_trans}
    p(\bd{x}_{t+1} \mid \bd{x}_{1:t}, \bth, \balpha) = \mathcal{N}\left(f(\bd{x}_{t-2:t}, \bth, \balpha), \sigma^2/R^2 \right)
\end{equation}
where
\begin{equation}\label{eq:BH_f}
    f\left(\bd{x}_{t-2:t}, \bth, \balpha\right) = \frac{1}{R}\sum_{j=1}^{J} \frac{\exp{\left[\beta \left(\bd{x}_t - R \bd{x}_{t-1}\right)\left(g_j \bd{x}_{t-2} + b_j - R \bd{x}_{t-1}\right)\right]}}{\sum_{j' = 1}^{J} \exp{\left[\beta \left(\bd{x}_t - R \bd{x}_{t-1}\right)\left(g_{j'} \bd{x}_{t-2} + b_{j'} - R \bd{x}_{t-1}\right)\right]}}\left(g_j \bd{x}_t + b_j\right)
\end{equation}
and $\balpha = (R, \beta, \sigma)$. The model is taken to be initialised with $\bd{x}_{-2} = \bd{x}_{-1} = \bd{x}_0 = 0$.

\subsection{The asset prices as differentiable and deterministic transformations of input noise}\label{app:BH_rep}

We claim in the main body that we may rewrite the $\bd{x}_t$ as deterministic transformations of standard Normal random variables. By exploiting the autoregressive structure of the model, we explicitly provide these forms for $\bd{x}_1$ and $\bd{x}_2$ below to demonstrate this claim. Throughout, $\epsilon_t \sim \mathcal{N}(0,1)$ are \textit{iid} standard Normal random variables. They are as follows:
\begin{align}
    \bd{x}_1 &= \frac{1}{R}\left(\sum_{j=1}^{J} \frac{b_j}{J} + \sigma\epsilon_1\right) := f_1(\epsilon_{1}, \bth, \balpha)\\
    \bd{x}_2 &= \frac{1}{R} \sum_{j = 1}^J \frac{\exp\left[{\frac{\beta b_j}{R} \left( \sum_{j'' = 1}^J \frac{b_{j''}}{J} + \sigma \epsilon_1 \right)}\right]}{\sum_{j' = 1}^J \exp\left[{\frac{\beta b_{j'}}{R} \left( \sum_{j'' = 1}^J \frac{b_{j''}}{J} + \sigma \epsilon_1 \right)}\right]} + \frac{\sigma}{R} \epsilon_2 := f_2(\epsilon_{1:2}, \bth, \balpha).
\end{align}

Repeating this process, we find that the $\bd{x}_t$ may all be expressed in the form $\bd{x}_t = f_t(\epsilon_{1:t}, \bth, \balpha)$ for a deterministic mapping $f_t : \mathbb{R}^{t} \times \mathbb{R}^{2J} \times \mathbb{R}^3 \to \mathbb{R}$.

Taking
\begin{align}
    \ell(\bd{y}, \bth) &= \text{MMD}^2(\mathbb{P}_T, \mathbb{P}_{\bth})\\
    &= \mathbb{E}_{x, x' \sim \mathbb{P}_{\bth}}\left[k(x, x')\right] + \mathbb{E}_{y, y' \sim \mathbb{P}_{T}}\left[k(y, y')\right] - 2\mathbb{E}_{x \sim \mathbb{P}_{\bth}, y\sim \mathbb{P}_{T}}\left[k(x, y)\right]\\
    &\approx \frac{1}{T(T-1)}\sum_{t\neq t'} k(\bd{x}_t, \bd{x}_{t'}) + \frac{1}{T(T-1)}\sum_{t\neq t'} k(\bd{y}_t, \bd{y}_{t'}) - \frac{2}{T^2}\sum_{t, t' = 1}^T k(\bd{x}_t, \bd{y}_{t'})
\end{align}
with a Gaussian kernel $k$, the loss $\ell(\bd{y}, \bth)$ is a deterministic and differentiable transformation of the noise drawn from the base distribution $\rho$ of the normalising flow and of the separate noise source with distribution $\nu$ given as input to the simulator. This permits us to estimate the gradient of the first term in \eqref{eq:gvi_obj} as
\begin{align}
    \nabla_{\phi} \mathbb{E}_{q_{\phi}}\left[ \ell(\bd{y}, \bth)\right] 
    &= \nabla_{\phi}\mathbb{E}_{u\sim \rho}\left[\ell(\bd{y}, \bth_{\phi}(u))\right]\\
    &= \mathbb{E}_{u\sim \rho}\left[\nabla_{\phi}\ell(\bd{y}, \bth_{\phi}(u))\right]\\\nonumber
    &= \mathbb{E}_{u\sim \rho}\left[\frac{1}{T(T-1)}\sum_{t\neq t'} \nabla_{\phi} k\left(f_t(\epsilon_{1:t}, \bth_{\phi}(u), \balpha), f_{t'}(\epsilon_{1:t'}, \bth_{\phi}(u), \balpha)\right) - \right.\\\label{eq:grad_reparam_bh}
    &\quad \quad \quad \quad \quad \quad \quad \quad \quad \quad \quad \quad \quad \quad \quad \quad \quad \quad \left. \frac{2}{T^2}\sum_{t, t' = 1}^T \nabla_{\phi} k(f_t(\epsilon_{1:t}, \bth_{\phi}(u), \balpha), \bd{y}_{t'})\right]
\end{align}

where in the first line we use the Law of the Unconscious Statistician and assume throughout that the order of derivatives and integrals can be exchanged freely.

\subsection{Further experimental results for the Brock \& Hommes model}\label{app:BH_exp}

\subsubsection{Calibration results with gradient horizon $H=0$}

\begin{figure}[h]
    \centering
    \includegraphics[width=0.5\columnwidth]{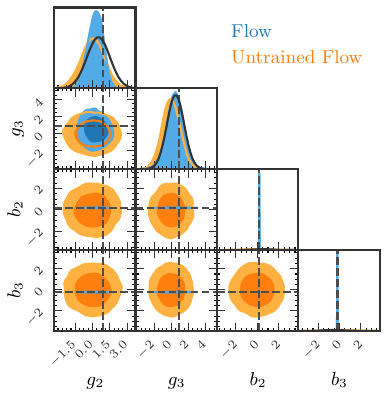}
    \caption{The inferred generalised posterior for the Brock \& Hommes model using a gradient horizon $H=0$.}
    \label{fig:bh_posterior}
\end{figure}

In Figure \ref{fig:bh_posterior}, we show the generalised posterior approximated by the converged normalising flow and with gradient horizon $H=0$, which achieved a loss close to 0. Since the objective function is lower-bounded by 0, this posterior can be taken to be a good approximation to the generalised posterior it targets.

\subsubsection{Further evidence in support of the bias-variance trade-off in the reparameterised Monte Carlo gradient estimator at different gradient horizons}

To further test the hypothesis that the Monte Carlo gradient estimators at different gradient horizons result in a bias-variance trade-off that can result in favourable performance at finite gradient horizons, we inspect the empirical distribution of the estimates $\eta_N$ of the gradient \eqref{eq:grad_reparam_bh} based on $N$ Monte Carlo samples,
\begin{equation}\label{eq:grad_N_bh}
    \eta_N := \frac{1}{N} \sum_{n=1}^{N} \left[
    \frac{1}{T(T-1)}\sum_{t\neq t'} \nabla_{\phi} k\left(\bd{x}_{t}, \bd{x}_{t'}
    \right) - \frac{2}{T^2}\sum_{t, t' = 1}^T \nabla_{\phi} k\left(
        \bd{x}_{t}, \bd{y}_{t'}
    \right)
    \right]
\end{equation}
where here we take a diagonal Gaussian distribution over $\mathbb{R}^4$ as the posterior estimator $q_{\phi}$. In this experiment, therefore, $\phi = (\mu_1, \ldots, \mu_4, \sigma_1, \ldots, \sigma_4)$, where $\mu_i$ and $\sigma_i$ are the mean and standard deviation in each dimension of this choice of $q_{\phi}$.

When implemented in the form given by Equations \eqref{eq:BH_trans} and \eqref{eq:BH_f} -- as is necessary to avoid the cumbersome task of finding the explicit form of $f_t(\epsilon_{1:t}, \bth, \balpha)$ for each $t$ -- the $\bd{x}_t$ depend on $\bth$ both explicitly and implicitly via $\bd{x}_{t-3:t-1}$. Thus in general we have

\begin{equation}\label{eq:grad}
    \frac{\partial \bd{x}_t}{\partial \bth_i} = \sum_{l = 1}^{\infty} \sum_{\substack{{\text{Length } l \text{ paths}} \\ {\mathbf{v} = (v_0, v_1, \ldots, v_{l-1}, v_l)} \\ \text{with } v_0 = \bd{x}_t, v_l = \bth_i}} {\prod_{i=0}^{l-1} \frac{\partial v_i}{\partial v_{i+1}} },
\end{equation}

where we abuse notation by taking the derivative on the left-hand side to mean ``holding only $(\bth_1, \ldots, \bth_{i-1}, \bth_{i+1}, \ldots, \bth_d)$ constant'' while the partial derivatives on the right-hand side are partial derivatives in the true sense of the term (or equivalently the $\bd{x}_t$ on the left-hand side is viewed only as a function of $\bth$, while on the right-hand side they are viewed as functions of both $\bth$ and $\bd{x}_{1:t}$). Choosing a gradient horizon of $H > 0$ then amounts to retaining paths of length greater than 1 if their first edge corresponds to an edge connecting $\bd{x}_t$ to any node in $\mathcal{X}_H := \{\bd{x}_{t-h} : h \in \mathcal{H}_H\}$, where $\mathcal{H}_0 = \emptyset$ and $\mathcal{H}_j = \{1, \ldots, j\}$ when $j > 0$. In this way, the derivative \eqref{eq:grad} is then taken instead as

\begin{equation}
    \frac{\partial \bd{x}_t}{\partial \bth_i} = \frac{\partial \bd{x}_t}{\partial \bth_i} + \sum_{l = 2}^{\infty} \sum_{\substack{{\text{Length } l \text{ paths}} \\ {\mathbf{v} = (v_0, v_1, \ldots, v_{l-1}, v_l)} \\ \text{with } v_0 = \bd{x}_t, v_1 \in \mathcal{X}_H, v_l = \bth_i}} {\prod_{i=0}^{l-1} \frac{\partial v_i}{\partial v_{i+1}} },
\end{equation}

where the same abuse of notation is once again used. This elimination of terms from the summation can be expected to reduce the variance since for two random variables $X_0, X_1$, it is the case that
\begin{equation}
    \text{Var}(X_0 + X_1) = \text{Var}(X_0) + \text{Var}(X_1) + 2 \text{Cov}(X_0, X_1),
\end{equation}
which can be greater than $\text{Var}(X_0)$ if $\text{Var}(X_1) + 2 \text{Cov}(X_0, X_1) > 0$. It may also be preferable to stricter truncation of the computation graph -- for example, by pruning all paths beyond a certain length -- as it retains information on long-range dependencies while still potentially reducing variance.

\begin{figure*}[h]
    \centering
    \includegraphics[width=\columnwidth]{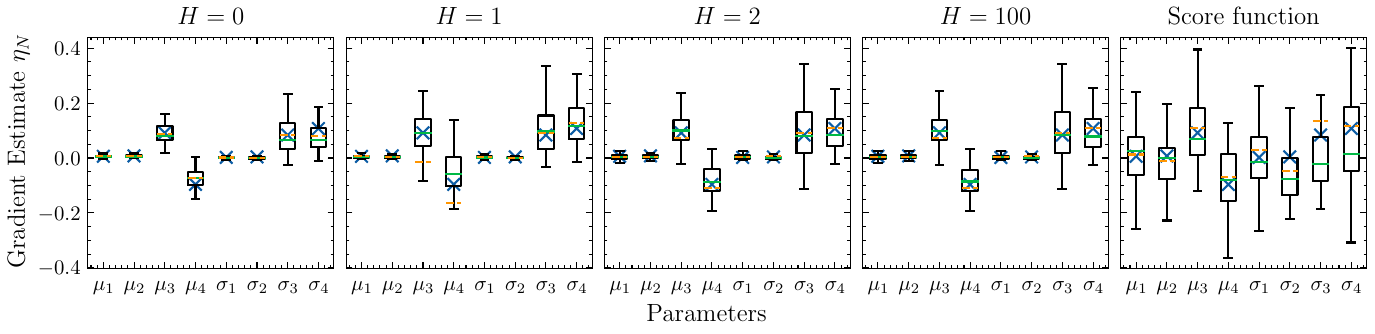}
    \vspace{-2em}
    \caption{Boxplots for the empirical distribution of Monte Carlo gradient estimates from gradient horizons $H=0,1,2,100$, and from the score-based gradient estimate, \eqref{eq:score_grad_bh_norm}. Orange (green) dashed lines show the mean (median) of the distributions. Blue crosses show the mean of the distribution of the gradient estimate \eqref{eq:score_grad_bh_norm} obtained using $N=1000$.}
    \label{fig:gradients_loss}
\end{figure*}

We plot in \autoref{fig:gradients_loss} boxplots for the distribution of $\eta_N$ obtained with $N=5$ and with different values of $H$, obtained at a fixed value for $\phi$ (the results were qualitatively similar for different $\phi$ we tried, and so we show only the results from one settings). We also show the same boxplots for the gradient estimate obtained with the score-based estimator,
\begin{equation}\label{eq:score_grad_bh_norm}
    \nabla_{\phi}\mathbb{E}_{q_{\phi}} \left[\ell(\bd{y}, \bth)\right] = \mathbb{E}_{q_{\phi}} \left[\ell(\bd{y}, \bth)\nabla_{\phi}\log q_{\phi}(\bth)\right] \approx \frac{1}{N} \sum_{n=1}^N \ell(\bd{y}, \bth^{(n)}) \nabla_{\phi} \log q_{\phi}(\bth^{(n)}).
\end{equation}

Orange (green) dashed lines show the mean (median) of the distributions. The blue crosses show the mean of the distribution of the gradient estimate \eqref{eq:score_grad_bh_norm} obtained using $N=1000$, which provide a good estimate of the target value (since the score-based estimator is unbiased). We see from this that, generally speaking, the variance of these estimates increase as $H$ increases, while the bias in the estimates do not degrade substantially. This highlights the possibility that using a finite 
gradient horizon can be beneficial when performing Monte Carlo gradient estimation for differentiable time series simulation models, such as ABMs, when reparameterisation is possible. Further work will be required to establish the general applicability and suitability of this technique.

\section{The JUNE model}
\label{app:june}

The JUNE model \cite{aylett-bullockJuneOpensourceIndividualbased2020} is an agent-based epidemiological model that generates a synthetic population at a highly detailed level using the English census data. This model has been applied in various scenarios, including analyzing the impact of the first and second waves of SARS-CoV2 in England \cite{vernonBayesianEmulationHistory2022c} and devising strategies to control disease transmission in refugee settlements \cite{aylett-bullockOperationalResponseSimulation2021c}.

To enhance its performance and enable gradient-based calibration, the JUNE model has been incorporated into the \textsc{GradABM} framework \cite{gradabm, queraDontSimulate}. This integration allows for faster execution and more efficient parameter calibration. The JUNE model offers a wide range of configurable parameters related to disease transmission and progression, vaccination, and non-pharmaceutical interventions.

Given a susceptible agent exposed to an infection at
location $L$, the probability of that agent getting infected is given b
\begin{equation}
    p = 1 - \exp{\left(-\psi_s \; \beta_L \; \Delta t\sum_{i\in g}\mathcal I_i(t) \right) },
\end{equation}
where the summation is conducted over all contacts an agent has with infected individuals at the given location $L$. The term $\mathcal I_i(t)$ represents the time-dependent infectious profile of each infected agent, while $\Delta t$ is the duration of the interaction. Additionally, $\beta_L$ corresponds to a location-specific parameter that captures the variation in the nature of interactions across different locations.

Since the $\beta_L$ parameters are not directly measurable physical quantities, they are typically calibrated using available data on the number of cases or fatalities over a specific time period. For the current work we consider the calibration of 11 $\beta_L$ parameters corresponding to the contact intensity at households, companies, schools, universities, pubs and restaurants, gyms, cinemas, shops, care homes, and residence visits. Additionally, we also calibrate the initial number of infections, $I_0$, which are distributed randomly across the population. The synthetic ground truth data is generated by using $I_0 = 10^{-3.5} N_a$ where $N_a$ is the number of agents, and $\beta_\mathrm{household} = \beta_\mathrm{care home} = 0.6$, $\beta_\mathrm{school} = \beta_\mathrm{company} = \beta_\mathrm{university} = 0.4$, $\beta_\mathrm{pub} =\beta_\mathrm{shop} =\beta_\mathrm{gym} =\beta_\mathrm{cinema} =\beta_\mathrm{visit} =0.1$.

The normalising flow is trained setting $\ell(\mathbf y, \bth)$ to be the squared distance between the $\log_{10}$ of the infection time-series. This choice of loss function keeps the training robust against outliers, since the number of infections can oscillate between several orders of magnitude. The specific training parameters are described in \autoref{app:training}. 

\subsection{Further experimental results for the JUNE model}\label{app:june_exp}

\begin{figure}[h]
    \centering
    \includegraphics[width=0.5\columnwidth]{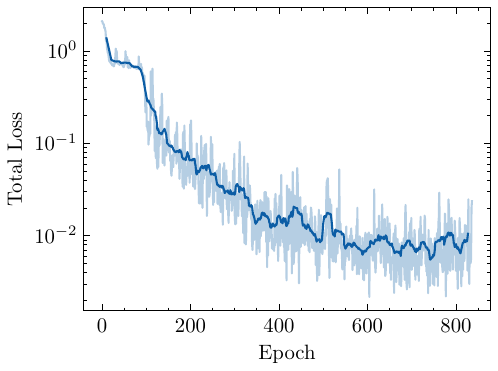}
    \caption{Loss function per epoch for the \june calibration.}
    \label{fig:june_loss}
\end{figure}

\begin{figure}[h]
    \centering
    \includegraphics[width=\columnwidth]{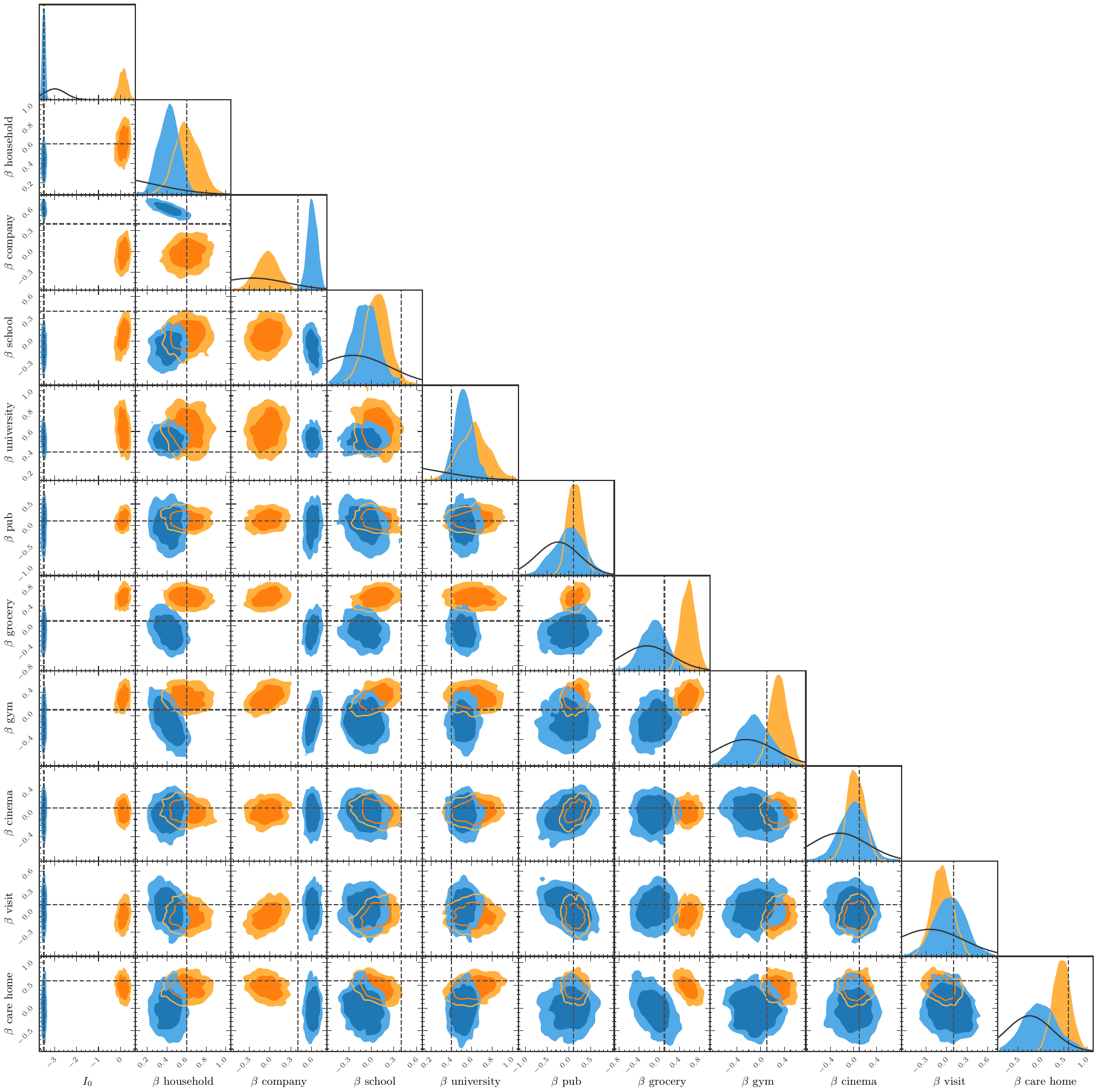}
    \caption{Corner plot of samples from the untrained normalising flow, the trained normalising flow, and the prior for the \june model calibration.}
    \label{fig:june_posterior}
\end{figure}

We show in \autoref{fig:june_loss} the loss as a function of epoch when performing SVI with \gradjune. We observe a rapid convergence after ~600 epochs. Since we are sampling 5 Monte Carlo samples to estimate \autoref{eq:grad_gen_obj}, this results in ~3,000 model evaluations. We also make a corner plot of the train and untrained normalising flow which we show in \autoref{fig:june_posterior}, where the solid black line denotes the prior density. We observe that the flow is very confident about the value of more sensitive parameters such as the initial number of infections and the contact intensity at companies, while it is less certain for venues which have a low impact in the overall number of infections, such as cinemas. It is worth noting that this calibration challenge is very underdetermined, that is, it is quite difficult by just observing the overall number of infections over time to infer the contact intensities at each location. Nonetheless, the flow fits well the synthetic ground truth data.

\section{Further experimental details}
\label{app:training}

We use the \textsc{normalizing-flows} library \cite{normflows} to implement the normalising flows in PyTorch. All models are trained using the AdamW optimizer \cite{Loshchilov2017DecoupledWD} with a learning rate of $10^{-3}$.

To calibrate the Brock \& Hommes and \june models, we employ a masked affine autoregressive flow \cite{NIPS2017_6c1da886} with 16 transformations, each parametrized by 2 blocks with 20 hidden units. We also set the regularisation weight to $w = 10^{-3}$ for both models and estimate \autoref{eq:grad_gen_obj} using 5 Monte Carlo samples.

\vspace{40ex}

\end{document}